\documentstyle[aps,prd,preprint,psfig]{revtex}
\begin{document}
\tightenlines

\def\stacksymbols #1#2#3#4{\def\theguybelow{#2}
    \def\verticalposition{\lower#3pt}
    \def\spacingwithinsymbol{\baselineskip0pt\lineskip#4pt}
    \mathrel{\mathpalette\intermediary#1}}
\def\intermediary#1#2{\verticalposition\vbox{\spacingwithinsymbol
      \everycr={}\tabskip0pt
      \halign{$\mathsurround0pt#1\hfil##\hfil$\crcr#2\crcr
               \theguybelow\crcr}}}
\def\lapproxeq{\stacksymbols{<}{\sim}{2.5}{.2}}
\def\gapproxeq{\stacksymbols{>}{\sim}{3}{.5}}

\title{The Wave Function Discord}
\author{Alexander Vilenkin\footnote{Electronic address: vilenkin@cosmos2.phy.tufts.edu}}

\address{Institute of Cosmology,
        Department of Physics and Astronomy,\\
        Tufts University,
        Medford, Massachusetts 02155, USA}
\date{\today}
\maketitle

\begin{abstract}

Linde's proposal of a Euclidean path integral with the ``wrong'' sign
of Euclidean action is often identified with the tunneling proposal
for the wave function of the universe.  However, the two proposals are
in fact quite different.  I illustrate the difference and point out
that recent criticism by Hawking and Turok does not apply to the
tunneling proposal.

\end{abstract}

\bigskip

The debate about the form of the wave function of the universe has
recently intensified \cite{1,2,3,4,5}. 
Its most recent round was initiated by Bousso and Hawking \cite{1} who
claimed that the tunneling proposal for the wave function of the
universe leads to a catastrophic instability of de Sitter space with
respect to pair-production of black holes. This claim was analyzed 
in Ref. \cite{2} and was shown to be unfounded. Linde \cite{3} has argued that in models of quantum creation of an open universe, the Hartle-Hawking wave function leads to unacceptably low values of the density parameter, and therefore the tunneling proposal should be preferred. In response, Hawking and Turok \cite{4} asserted that with the choice of the tunneling wave function, all perturbations about a homogeneous cosmological background become unstable and therefore this wave function can be meaningfully defined only in homogeneous minisuperspace models. This led them to conclude that the Hartle-Hawking wave function is the only proposal "with some pretentious to completeness" \cite{5}.
     
While the level of rhetoric is high, I think the progress on this issue may be helped by pointing out that what goes under the name of "tunneling wave function" are in fact two completely different wave functions. The debate will certainly gain in clarity if this fact is fully recognized.
     
Let me now briefly review the different proposals. The Hartle-Hawking wave function is given by the integral \cite{6}
\begin{equation}
\psi_{HH} = \int e^{-S_E},
\label{1}
\end {equation}
where $S_E$ is the Euclidean action and the integration is taken over compact Euclidean geometries and matter fields with a specified field configuration at the boundary. The Euclideanization is achieved by the standard Wick rotation, $t\to -i\tau$, from a Lorentzian path integral.
Linde \cite{7} suggested that the Euclidean rotation should be
performed in the opposite sense, yielding
\begin{equation}
\psi_L = \int e^{+S_E}.
\label{2}
\end{equation}
Finally, I introduced the tunneling wave function $\psi_T$ which is
specified either by the tunneling boundary condition \cite{8} or by a
Lorentzian path integral, 
\begin{equation}
\psi_T = \int e^{iS},
\label{3}
\end{equation}
interpolating between "nothing" and a specified field configuration
\cite{9}.
Arguments that the two definitions are equivalent were presented in \cite{10}.

Hawking and Turok's criticism \cite{4} is directed against Linde's
wave function (\ref{2}). They point out that while the anti-Wick
rotation $t\to +i\tau$ may work in simple minisuperspace models, it
leads to disastrous consequences when inhomogeneous modes are
included. All such modes become unstable, resulting in a breakdown of
the semiclassical description of the universe. Linde has argued that
this problem might be avoided if the Euclidean rotation is performed
separately for the background fields and for the perturbations
\cite{3}, but no prescription of this sort has yet been suggested that
would apply in the general case.
On the other hand, for the tunneling wave function it was shown in \cite{VV} that the gravitational and matter fields are stable and are in the same quantum state as for the Hartle-Hawking wave function.
     
If the reader needs further convincing that the tunneling and Linde's
wave functions are indeed different, I would like to illustrate the
difference for the simplest de Sitter minisuperspace model where both wave
functions are well behaved. In this model, the universe is assumed to
be homogeneous, isotropic, closed, and filled with a vacuum of
constant energy density $\rho_v$ \cite{F2}. The radius of the universe $a$ is the only variable of the model, and the wave function $\psi(a)$ satisfies the Wheeler-DeWitt equation
\begin{equation}
\left[{d^2 \over{da^2}} - a^2(1 - H^2a^2)\right]\psi(a) = 0.
\label{WDW}
\end{equation}
Here, $H^2 = 8\pi G\rho_v/3$ and I have disregarded the ambiguity in the ordering of non-commuting operators $a$ and $d/da$. (This ambiguity is unimportant in the semiclassical approximation which I am going to use below).

Eq.(\ref{WDW}) has the form of a one-dimensional Schrodinger
equation for a `particle' described by a coordinate $a(t)$, having
zero energy, and moving in a potential 
\begin{equation}
U(a)=a^2(1-H^2a^2).  
\end{equation}
The classically
allowed region is $a\geq H^{-1}$, and the WKB solutions of
(\ref{WDW}) in this region are
\begin{equation}
\psi_\pm (a)=[p(a)]^{-1/2} \exp \left[ \pm i\int_{H^{-1}}^a p(a')da'
\mp i\pi /4 \right],
\label{psi1}
\end{equation}
where $p(a)=[-U(a)]^{1/2}$.  The under-barrier, $a<H^{-1}$, solutions
are
\begin{equation}
{\tilde \psi}_\pm
(a)=|p(a)|^{-1/2}\exp\left[\pm\int_a^{H^{-1}}|p(a')|da'\right].
\label{psi2}
\end{equation}

The classical momentum conjugate to $a$ is $p_a=-a{\dot a}$.
For $a\gg H^{-1}$, we have
\begin{equation}
(-id/da)\psi_\pm (a)\approx\pm p(a)\psi_\pm (a),
\end{equation}
and thus $\psi_-(a)$ and $\psi_+(a)$
describe an expanding and a contracting universe, respectively.  
The tunneling boundary condition requires that only the expanding
component should be present at large $a$,
\begin{equation}
\psi_T (a>H^{-1})=\psi_-(a).
\label{tunnel1}
\end{equation}
The under-barrier wave function is found from the WKB connection
formula,
\begin{equation}
\psi_T (a<H^{-1})={\tilde \psi}_+(a)-{i\over{2}}{\tilde\psi}_-(a).
\label{tunnel2}
\end{equation}
The growing exponential ${\tilde\psi}_-(a)$ and the decreasing
exponential ${\tilde\psi}_+(a)$ have comparable amplitudes at the
nucleation point $a=H^{-1}$, but away from that point the decreasing
exponential dominates (see Fig.~1).

The Hartle-Hawking wave function is specified by requiring that it is
given by $\exp (-S_E)$ in the Euclidean under-barrier regime.  This
gives \cite{6} 
\begin{equation}
\psi_H(a<H^{-1})={\tilde \psi}_-(a),
\end{equation}
\begin{equation}
\psi_H(a>H^{-1})=\psi_+(a)-\psi_-(a).
\end{equation}
Linde's wave function is obtained by reversing the sign of the
exponential in the Euclidean regime,
\begin{equation}
\psi_L(a<H^{-1})={\tilde \psi}_+(a),
\end{equation}
and the continuation to the classically allowed range of $a$ gives
\begin{equation}
\psi_L(a>H^{-1})={1\over{2}}[\psi_+(a)+\psi_-(a)].
\end{equation}
The Hartle-Hawking and Linde's wave functions are schematically
represented in Fig.~2.

We can now compare Linde's and tunneling wave functions.  Under the
barrier, both of them are dominated by the decaying exponential
${\tilde\psi}_+(a)$.  For this reason, $\psi_L$ and $\psi_T$ give 
similar expressions for the probability of the creation of the universe
\cite{3,9}.  But that is where the similarity ends.  The
sub-dominant growing exponential, which is absent in Linde's wave
function, was crucial in the derivation of the quantum state of
inhomogeneous modes for the tunneling wave function \cite{VV}.
Moreover, outside
the barrier, Linde's wave
function includes expanding and contracting universe components with equal
amplitudes, and in this respect it is more similar to Hartle-Hawking
than to tunneling wave function.  

In conclusion, Linde's wave function $\psi_L$ and tunneling wave
function $\psi_T$ are two different wave functions. Using the
term "tunneling wave function" for $\psi_L$ and $\psi_T$
indiscriminately, as it was done, e.g., in Refs. \cite{1,3}, has
introduced much confusion in the debate. The recent objection
raised by Hawking and Turok \cite{4} applies to $\psi_L$ but not
to $\psi_T$, so the tunneling wave function remains a viable
choice for the wave function of the universe.

I would like to add that, in following this wave function debate, the
reader should be aware that all three wave functions are far from
being rigorously defined mathematical objects.  Except in simplest
models, the actual calculations of these wave functions involve
additional assumptions which may appear reasonable, but are not really
well justified.  For a recent discussion of problems associated with
defining and interpreting the cosmological wave function see, e.g.,
Ref. \cite{10}.

\begin{figure}
\psfig{file=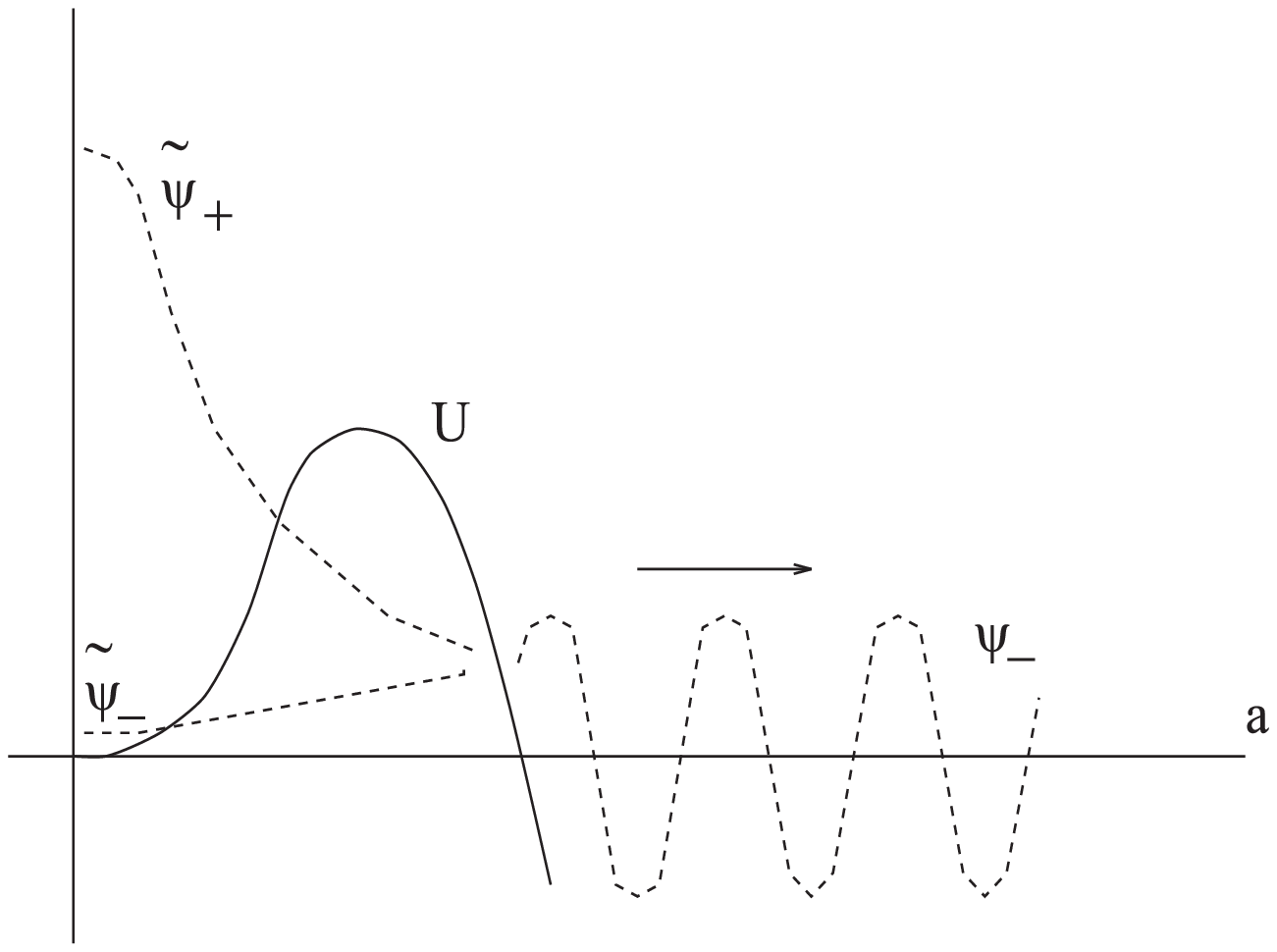}
\vspace{1.0in}
\caption{
Tunneling wave function for the de Sitter minisuperspace model.  The
potential $U(a)$ is shown by a solid line and the wave function by a
dashed line.
}
\end{figure}

\begin{figure}
\psfig{file=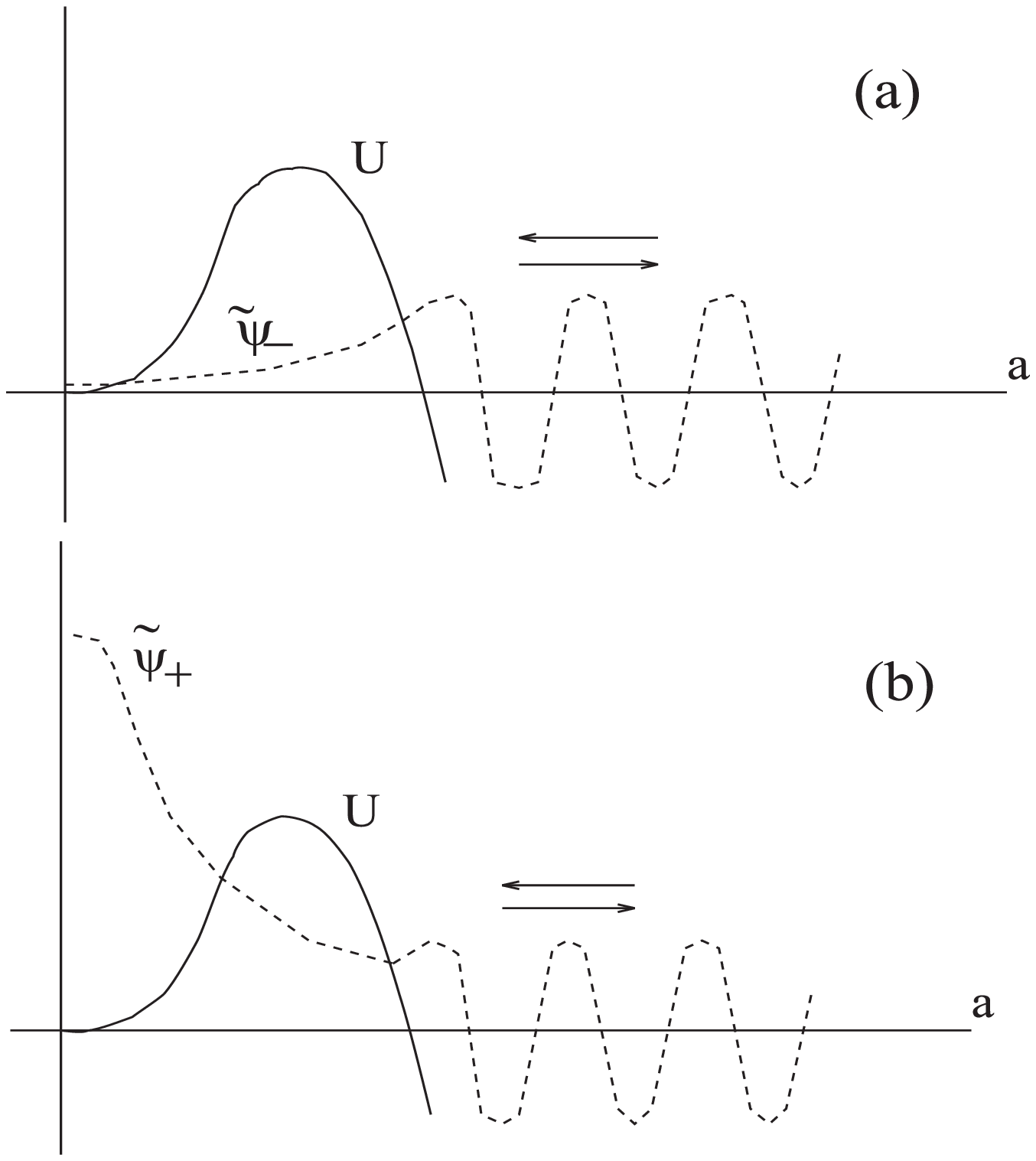}
\vspace{1.0in}
\caption{
Hartle-Hawking (a) and Linde (b) wave functions for de Sitter
minisuperspace model.  
}
\end{figure}


\begin{references}
\bibitem{F1}
In my first papers on quantum cosmology \cite{11} I used the standard
Wick rotation, but later I realized \cite{9} that it did not
correspond to the picture of ``tunneling from nothing'' which I had in
mind from the beginning.
\end{references}
\end{document}